\documentclass[sigconf]{acmart}

\AtBeginDocument{%
  \providecommand\BibTeX{{%
    \normalfont B\kern-0.5em{\scshape i\kern-0.25em b}\kern-0.8em\TeX}}}

\usepackage{amsmath,amsfonts}
\usepackage{graphicx}
\graphicspath{{Figure/}}
\usepackage{textcomp}
\usepackage{xcolor}
\usepackage[caption=false,font=normalsize,labelfont=sf,textfont=sf]{subfig}
\usepackage{url}
\usepackage{booktabs}
\usepackage{tabularx}
\usepackage{algorithm}
\usepackage{makecell}
\usepackage[noend]{algpseudocode}
\usepackage{enumitem}
\usepackage{multirow}
\usepackage{array}
\usepackage{multicol}
\usepackage{tcolorbox}
\usepackage{pdfpages}
\usepackage{float}
\usepackage{placeins}
\usepackage{soul}
\usepackage{listings}
\usepackage{adjustbox}
\usepackage{xfrac}
\usepackage[mathcal]{euscript}
\usepackage{pgfplots}
\usepackage{pgfplotstable}
\usepackage{tikz}
\usepackage{etoolbox}

\begin{document}
\pagestyle{empty}

\title{TAMI-MPC:\underline{T}rusted \underline{A}cceleration of \underline{M}inimal-\underline{I}nteraction \underline{MPC} for Efficient Nonlinear Inference}

\author{
Zhuoran Li$^{\dagger *}$, 
Hanieh Totonchi Asl$^{\dagger *}$, 
Yifei Cai$^{\ddagger}$, 
Ebrahim Nouri$^{\dagger}$, 
Danella Zhao$^{\dagger}$\\[4pt]
\small
$^{\dagger}$Department of Electrical and Computer Engineering, University of Arizona, Tucson, United States\\
\{zli1122, haniehta, ebinouri, danellazhao\}@arizona.edu\\
$^{\ddagger}$Department of Computer Science, Iowa State University, Ames, United States\\
yifeic@iastate.edu\\[4pt]
}

\begin{abstract}

Secure multi-party computation (MPC) offers a practical foundation for privacy-preserving machine learning at the edge. However, current MPC systems rely heavily on communication and computation-intensive primitives—such as secure comparison-for nonlinear inference, which are often impractical on resource-constrained platforms. 
To enable real-time inference under resource-constrained platform, we introduce a \underline{T}rusted \underline{A}cceleration of \underline{M}inimal-\underline{I}nteraction \underline{MPC} framework, TAMI-MPC, for nonlinear evaluation. Specifically, we reduce communication cost by redesigning the core primitives, leaf comparison and tree merge, reducing the interactive round from $\log_2 n$ to just $1$ per operation. Furthermore, unlike prior work that heavily relies on oblivious transfer (OT), a well-known computational bottleneck, we leverage synchronized seeds inside the TEE to eliminate OT for the vast majority of our designs, along with a correlated-randomness reuse technique that keeps new designs computationally lightweight. To fully realize the potential, we design a specialized accelerator that restructures the dataflow across stages to enable continuous, fine-grained streaming and high parallelism, reducing memory overhead. Our design achieves up to \(4.86\times\) speedup on ResNet-50 inference, compared with state-of-the-art CNN frameworks, and achieves up to \(7.44\times\) speedup on BERT-base inference, compared with state-of-the-art LLM frameworks.

\end{abstract}


\maketitle

\section{Introduction}

To address privacy concerns in machine learning as a service (MLaaS), privacy-preserving MLaaS (PPMLaaS) employs cryptographic primitives to protect sensitive data~\cite{juvekar2018gazelle, riazi2019xonn, rouhani2018deepsecure, mishra2020delphi, zhang2021gala}. Among these primitives, multiparty computation (MPC) has gained adoption for its effectiveness in handling nonlinear operations~\cite{rathee2020cryptflow2, huang2022cheetah, pang2024bolt, lu2025bumblebee, zhang2024individual}. 
A key module in this context is secure comparison, as many nonlinear functions such as ReLU, GELU, Max-pooling, and Softmax, ultimately reduce to comparison operations. Modern PPMLaaS systems~\cite{huang2022cheetah, pang2024bolt, lu2025bumblebee,feng2025panther} widely adopt the Millionaires’ protocol from Cryptflow2~\cite{rathee2020cryptflow2}, as it provides the state-of-the-art balance of low communication, high throughput, and seamless compatibility with secret sharing, making it particularly well-suited for large-scale deep learning inference.

\begin{figure}[t]
  \centering
  \includegraphics[width=\linewidth]{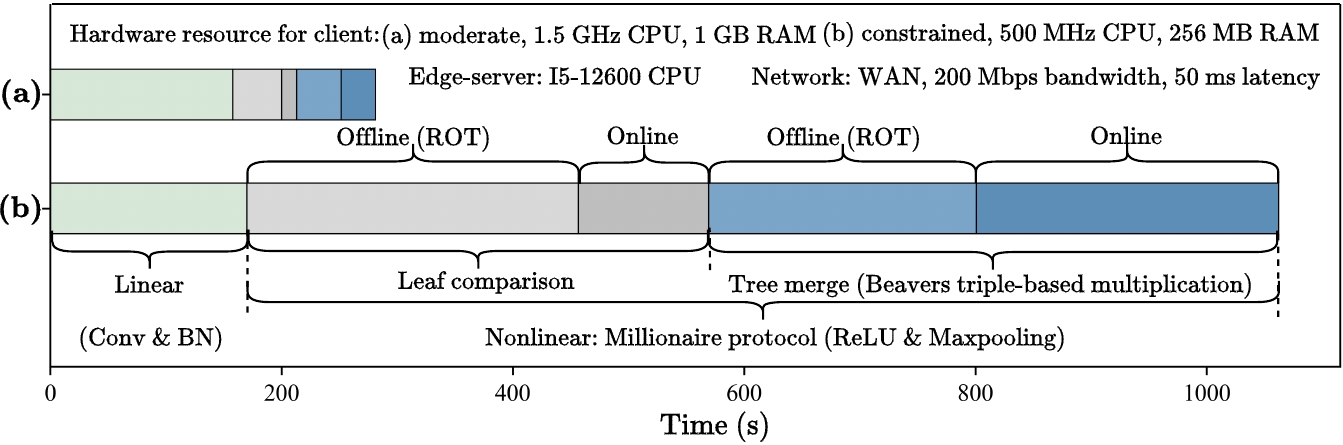}
  \caption{ResNet-50 inference performance with Cheetah~\cite{huang2022cheetah}.
  }
  \label{preliminary}
\end{figure}

\textbf{Motivations.} While the cost of the Millionaires' protocol is often moderate under fast network conditions and on consumer-grade hardware (e.g., smartphone or desktop-class CPUs), it becomes a major bottleneck in real-world inference with resource-constrained networks and client devices (e.g., IoT sensors, wearables, or portable medical devices with limited memory~\cite{zheng2019challenges,aminifar2024privacy}). Figure~\ref{preliminary} presents a time breakdown for state-of-the-art ResNet-50 inference under two resource settings: (i) moderate and (ii) constrained. The nonlinear phase incurs about a $9\times$ slowdown in the constrained setting. This overhead is driven primarily by the communication and computation costs of the two core operations: OT-based leaf comparison and tree merge via Beaver-triple–based multiplication.  

\textbf{Communication Overhead.} First, although OT-based leaf comparison and tree merge via Beaver’s triple–based multiplication offer a favorable trade-off in state-of-the-art PPMLaaS frameworks, their communication costs (data volume and number of interactive rounds) remain substantial. These costs are often acceptable on high-bandwidth, low-latency LANs, but they become a dominant bottleneck on constrained networks (e.g., limited bandwidth and high latency)~\cite{ryffel2022ariann,huang2024efficient,maeng2024accelerating}. Second, on resource-constrained clients, insufficient memory precludes pre-generating and storing the large volumes of correlated randomness required by these protocols. As a result, inference must be partitioned into small batches, and correlated randomness is produced just in time, which further increases the number of interactive rounds, leading to significant slowdowns in nonlinear evaluation.  

\textbf{Computation Overhead.} First, the two core operations, leaf comparison and tree merge, heavily rely on pre-generated correlated randomness (e.g., random oblivious transfer (ROT)), which has been reported as a computational bottleneck—exacerbated on resource-constrained devices—due to cache pressure~\cite{li2025silentflow,lin2025ironman} and limited opportunity for parallel execution. Second, in resource-constrained environments, the data movement and memory access patterns induce imbalanced dataflow, underutilize available compute resources, and saturate memory bandwidth due to the 1-bit nature of the tree-merge Boolean operations, which are executed on byte-addressable CPUs, forcing each bit to be stored and accessed as a full byte. As a result, system performance becomes constrained by memory throughput rather than raw computational capability.

The contributions are summarized as:

\begin{itemize}

    \item We present, Trusted Acceleration of Minimal-Interaction MPC, namely TAMI-MPC, a framework that (i) yields an $8\times$ reduction in communication cost (in bits) and reduces the interactive round complexity of leaf comparison and tree merge from $2$ and $\log_2 n$ to $1$, respectively (§\ref{Sec:complexity_sec}), and (ii) eliminates all interaction for offline correlated-randomness generation by deriving it from synchronized TEE seeds with a security-preserving randomness-reuse optimization, which avoids batch-by-batch communication under constrained clients and reduces TEE-side generation cost by over $79\times$.

    \item We design a specialized accelerator to further boost nonlinear evaluation, which (i) addresses the well-known ROT-generation computation overhead in leaf comparison via a \textit{pipeline-aware interleaving} of core operations (e.g., key expansion and AES encryption), improving efficiency by $3.97\times$, and (ii) reduces data-movement overhead during tree merge through a \textit{memory-efficient data-management scheme} that exploits parallelism via \textit{packed polynomial execution}, achieving a speedup of $80.15\times$.

\end{itemize}

\section{Background and Security Definition}\label{Sec:Preliminaries}

\subsection{Secure Comparison (Millionaires' Protocol)}

\begin{figure}[h]
  \centering
  \includegraphics[width=0.9\linewidth, trim={0cm 0cm 0cm 0cm}, clip]{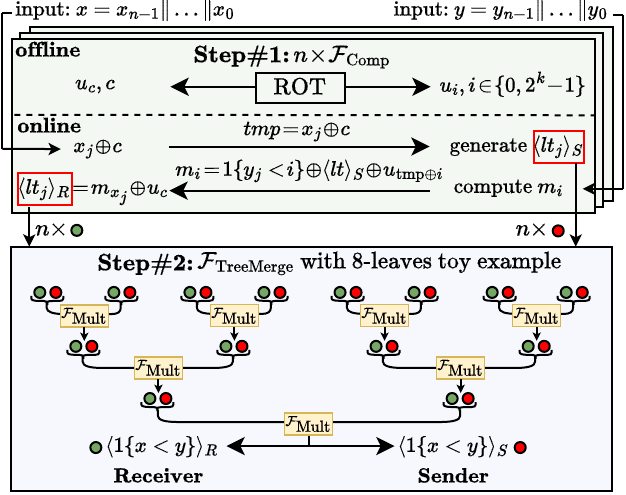}
  \caption{Illustration of Millionaires' protocol.}
  \label{millionaire_protocol}
\end{figure}

Prior works such as CryptFlow2~\cite{rathee2020cryptflow2} employ the Millionaires' protocol for secure comparison, which has become a standard primitive in state-of-the-art PPMLaaS frameworks~\cite{huang2022cheetah, pang2024bolt,lu2025bumblebee}. The Millionaires' protocol consists of two key steps: (1) OT-based leaf comparisons, \(\mathcal{F}_{\text{Comp}}\), over smaller input blocks, and (2) a tree merge phase, \(\mathcal{F}_{\text{TreeMerge}}\), implemented via Beaver’s triple–based secure multiplications \(\mathcal{F}_{\text{Mult}}\) to obtain the final secure comparison result, as shown in Figure~\ref{millionaire_protocol}.

In $\mathcal{F}_{\text{Comp}}$ (Step~\#1), an offline phase prepares correlated randomness via ROT to mask the online OT messages. This preprocessing consumes $k\times$ ROT instances ($k$ is input bitlength of $y_j$) and incurs computation and communication overhead~\cite{lin2025ironman, li2025silentflow}, while the online phase completes in two rounds. Specifically, the online input $x_j$ is masked with a pre-generated selection bit $c$ as $tmp = x_j \oplus c$, and $tmp$ is sent from the receiver to the sender. Given its input $y_j$, the sender prepares oblivious messages $m_i = \mathbf{1}\{y_j < i\} \oplus \langle lt \rangle_S\oplus u_{tmp \oplus i}$, for all $i \in \{0,2^k-1\}$. Messages $m_i$ contain the leaf comparison result of the sender's input $y_j$ and the receiver's input $x_j$, but is masked with the ROT-generated randomness $u_{tmp \oplus i}$, which is indexed via the masked choice $tmp$. Therefore, by the mechanics of OT, in the online phase the receiver can use $u_c$ to decrypt only the message corresponding to its selection choice $c$, and in this way it obtains the matching share $\langle lt \rangle_R$.

To avoid the exponential blow-up of comparing wide integers directly, for example, \(2^{32}\) OT messages for 32-bit, each input is partitioned into \(8\!\times\!4\)-bit chunks, so each 4-bit comparison needs only \(2^4\) online OT messages during the leaf comparison phase (Step~\#1). However, this blockwise merging design introduces an expensive communication cost: the per-block comparison bits must be securely multiplied layer by layer via Beaver’s triple–based multiplication to produce the final result (Step\#2, \(\mathcal{F}_{\text{TreeMerge}}\)). Each merge point requires \(2\times\!\) Beaver’s triple-based secure multiplications. Under the standard ROT-to-triple construction, this consumes \(4\times\!\) ROTs per merge. Therefore, the merging step contributes \(4(n-1)\) additional ROTs on top of the \(n\times k\) ROTs already incurred by \(n\times\mathcal{F}_{\text{Comp}}\). As observed in~\cite{li2025silentflow, lin2025ironman}, ROT generation dominates both computation and communication in practice, making it the primary bottleneck across both steps of the protocol. Moreover, even when all branches execute in parallel, the online communication depth remains \(\log_2 n\) for \(n\) chunks, introducing \(\log_2 n\) round-trip latency barriers that constitute another major bottleneck.

\subsection{Security Definition}

\begin{figure}[t]
  \centering
  \includegraphics[width=\linewidth, trim={0cm 0cm 0cm 0cm}, clip]{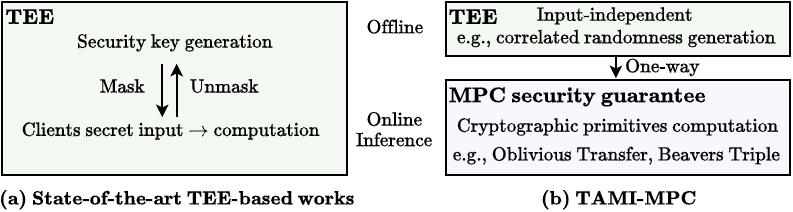}
  \caption{Comparison of security boundary between state-of-the-art TEE-based works~\cite{zhou2022ppmlac, zhou2022efficient} and TAMI-MPC.}
  \label{security_definition}
\end{figure}

The key difference between TAMI-MPC and previous TEE-based MPC systems lies in the security boundary as shown in Figure~\ref{security_definition}. Prior work~\cite{zhou2022ppmlac, zhou2022efficient} imports secret inputs (e.g., client data) into the TEE for online computation, requiring the TEE to protect these secrets during inference. In contrast, TAMI-MPC derives its inference security from MPC rather than the TEE: the TEE is restricted to offline phase input-independent tasks (e.g., correlated randomness generation), while inference is performed entirely via MPC primitives in the untrusted domain. Thus, the TEE is not involved during online inference; even if the TEE is compromised at this stage, both client and server secrets remain protected by MPC.  

\section{TAMI-MPC Secure Comparison $\mathcal{F}_{\text{Mill}}$}\label{Sec:Design}

In this section, we present the two core primitives—TEE-assisted leaf comparison $\mathcal{F}_{\text{Comp}}$ and tree merge $\mathcal{F}_{\text{PolyMult}}$—which together enable two-round (including offline and online) secure comparison (Millionaires' protocol, denoted as $\mathcal{F}_{\text{Mill}}$). We then further optimize this design by interleaving the two primitives and reusing correlated randomness, substantially improving the performance.

\subsection{TEE-assisted Leaf Comparison $\mathcal{F}_{\text{Comp}}$}\label{sec:comparison}

\begin{figure}[t]
  \centering
  \includegraphics[width=0.85\linewidth, trim={0cm 0cm 0cm 0cm}, clip]{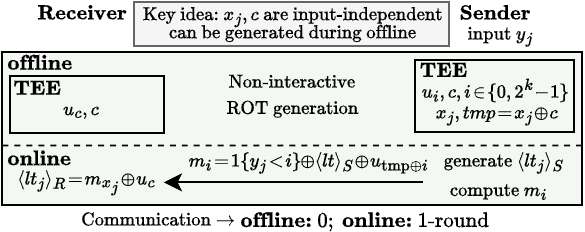}
  \vspace{0.03in}
  \caption{Design of $\mathcal{F}_{\text{Comp}}$}
  \label{TEE_secure_comparison}
\end{figure}

As shown in Figure~\ref{millionaire_protocol}, state-of-the-art leaf comparison builds on 1-out-of-2 OT and requires two online rounds. In the first round, the receiver sends $x_j \oplus c$ to the sender as the OT selection mask. This round can be eliminated if both parties locally derive $x_j$ and the selection bit $c$ from synchronized seeds via a pseudorandom generator (PRG) inside their TEEs. Our key observation is that, in OT-based leaf comparison for nonlinear evaluation, one of the two OT inputs is always an input-independent random value~\cite{rathee2020cryptflow2, huang2022cheetah, pang2024bolt, lu2025bumblebee}. Concretely, after a linear layer computation (e.g., a convolution), the server returns a homomorphically encrypted output $\mathrm{Enc}(output)$ masked by a one-time mask $M$, so the client receives $\mathrm{Enc}(output) - mask$. This masking protects the server’s secrets (e.g., model weights). The next step runs the Millionaires’ protocol on $(output - mask,\, mask)$. Because $mask$ is independent of the client input and can be derived within the TEE, these masks can be generated offline without communication while preserving security. In our leaf comparison, the input $x$ equals the $mask$; hence $x_j$ (Figure~\ref{TEE_secure_comparison}) can be deterministically generated on both sides via synchronized PRG within TEE during the offline phase.

The remaining challenge is the selection bit \(c\): in standard protocols, \(c\) is produced during ROT generation and revealed only to the receiver. Building on SilentFlow~\cite{li2025silentflow}, which derives ROTs non-interactively inside the TEE, we observe that \(c\) appears as an internal intermediate that is normally discarded. Our key idea is to retain \(c\) inside the TEE at both parties (never exposing it to the host), so each side deterministically derives the same structured randomness for \(c\) from synchronized seeds. Combined with locally deriving the input-independent mask \(x_j\), this removes the first online round in Figure~\ref{millionaire_protocol}.

It should be noted that $x_j$ and $c$ are input-independent, yet retaining them in the TEE poses a security risk if the TEE is compromised. For example, knowing $x_j$ could reveal the receiver’s parameters (e.g., model weights), while knowing $c$ could expose related OT values. In our design, only $tmp$ is needed during the online phase for secure comparison; $x_j$ and $c$ are discarded after computing $tmp$. This ensures the protocol remains secure without relying on TEE protection during the \textit{online phase}, aligning with our goal of maintaining \textit{online phase} security even if the TEE is breached—a key distinction from prior TEE-based approaches~\cite{zhou2022ppmlac,huang2024efficient}.

\subsection{TEE-assisted Tree Merge $\mathcal{F}_{\text{PolyMult}}$}\label{sec:tree-merging}

\begin{figure}[t]
  \centering
  \includegraphics[width=\linewidth, trim={0cm 0cm 0cm 0cm}, clip]{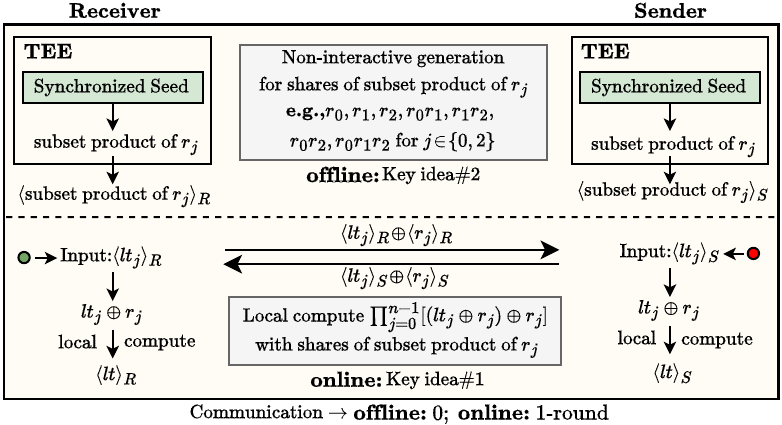}
  \caption{Design of $\mathcal{F}_{\text{PolyMult}}$}
  \label{TEE_secure_multiplication}
\end{figure}

State-of-the-art tree merge operation (\(\mathcal{F}_{\text{TreeMerge}}\)) can be viewed as a sequence of multiplications. In current methods, each level must wait for the previous level’s outputs, leading to \(\log_2 n\) online rounds (Figure~\ref{millionaire_protocol}) and intensive offline work to prepare Beaver triples. To remove this sequential dependency, we perform a one-round polynomial multiplication that merges all \(n\) inputs at once as shown in Figure~\ref{TEE_secure_multiplication}. Specifically, we rewrite the product in masked form and expand it under the additive secret sharing form:
\begin{equation}
\prod_{j=0}^{n-1} lt_j
\!=\!
\prod_{j=0}^{n-1}\big[lt_j\oplus r_j\oplus r_j\big]
\!=\!\prod_{j=0}^{n-1}(\sum_{s_j=0}^{1} (lt_j\oplus r_j)^{s_j}\,r_j^{\,1\oplus s_j}),
\end{equation}
expanding this expression gives every term as a subset-product of \(lt_j-r_j\) factors and a subset-product of the \(r_j\)’s. The only messages that need to be exchanged online are the masked differences \(lt_j-r_j\). Each party obtains these by exchanging their additive shares \(\langle lt_j\rangle\!\oplus\!\langle r_j\rangle\) and reconstructing \(lt_j\oplus r_j\). After this one exchange, each side completes its output share locally using pre-shared subset-product shares of the \(r_j\)’s. 

One challenge is that such subset-product shares of \(r_j\) are typically generated with communication-heavy cryptographic primitives such as ROT. Instead, we derive them non-interactively from synchronized seeds inside the TEE, eliminating the reported ROT generation computation and communication overhead~\cite{rathee2020cryptflow2,huang2022cheetah,pang2024bolt,lu2025bumblebee}. Consequently, the \(\log_2 n\) online rounds plus \(4\!\times\!(n-1)\;\text{ROTs generation}\) collapse to a single round that exchanges the \(lt_j\oplus r_j\) values; all remaining work is local. 

We use the product of \(n=3\) terms as an illustrative example to explain the one-round polynomial multiplication; the actual tree-merge formulation based on our one-round polynomial multiplication is presented in §\ref{sec:fusion}. Specifically, consider the one-round polynomial multiplication for \(n=3\). Define \(\tilde{lt}_0 = lt_0 \oplus r_0\), \(\tilde{lt}_1 = lt_1 \oplus r_1\), and \(\tilde{lt}_2 = lt_2 \oplus r_2\). Let \(p \in \{0,1\}\) denote the party (\(p=0\) for the client and \(p=1\) for the server). The secure product can then be expanded and evaluated locally as follows:

\vspace{-0.1in}
\begin{align}
{\textstyle\prod_{j=0}^{2}lt_j}
&= (\tilde{lt}_0\oplus r_0)(\tilde{lt}_1\oplus r_1)(\tilde{lt}_2\oplus r_2) \label{eq:prod-lt}\\
&= \tilde{lt_0}\tilde{lt_0}\tilde{lt_0}
 \oplus r_0\,\tilde{lt_1}\tilde{lt_2}
 \oplus r_1\,\tilde{lt_0}\tilde{lt_2}
 \oplus r_2\,\tilde{lt_0}\tilde{lt_1} \nonumber \\ 
&\oplus r_0r_1\,\tilde{lt_2}
 \oplus r_0r_2\,\tilde{lt_1}
 \oplus r_1r_2\,\tilde{lt_0}
 \oplus r_0r_1r_2. \nonumber
\end{align}

\noindent The parties compute local shares as
\vspace{0.01in}
\begin{align}
\langle {\textstyle\prod\nolimits_{j=0}^{2}}lt_j \rangle_{p} &=
\langle r_0\rangle_p\,\tilde{lt_1}\tilde{lt_2}
\oplus \langle r_1\rangle_p\,\tilde{lt_0}\tilde{lt_2}
\oplus \langle r_2\rangle_p\,\tilde{lt_0}\tilde{lt_1} \label{eq:share-R}\\
&\oplus \langle r_0r_1\rangle_p\,\tilde{lt_2}
\oplus \langle r_0r_2\rangle_p\,\tilde{lt_1}
\oplus \langle r_1r_2\rangle_p\,\tilde{lt_0}
\oplus p\langle r_0r_1r_2\rangle_p. \nonumber
\end{align}


\subsection{Further Optimization}\label{sec:fusion}

\noindent\textbf{Opt.\#1: Interleaving leaf comparison and tree-merge to reduce half of the communication.} Because the inputs of tree merge multiplication $\mathcal{F}_{\text{PolyMult}}$ are produced by the leaf comparison, we note that on one side (e.g., the sender in Figure~\ref{TEE_secure_comparison}), the output $\langle lt_j\rangle_S$ is an input-independent, locally generated share (Algorithm 1 in~\cite{rathee2020cryptflow2}). Thus, we move this input-generation step into the TEE, enabling both parties to derive it via synchronized seeds during the offline phase. In the subsequent $\mathcal{F}_{\text{PolyMult}}$, the receiver no longer needs to obtain \(\langle lt_j\rangle_S \oplus \langle r_j\rangle_S\) from the sender; it is already available from the leaf comparison step inside the TEE. The masked difference can then be formed locally as \(lt_j \oplus r_j=(\langle lt_j\rangle_R \oplus \langle r_j\rangle_R) \oplus (\langle lt_j\rangle_S \oplus \langle r_j\rangle_S)\), with the sender’s term released from the TEE. This removes one of the two cross-party messages in the $\mathcal{F}_{\text{PolyMult}}$, effectively halving its communication cost.

\noindent\textbf{Opt.\#2: Reusing correlated randomness during} $\mathcal{F}_{\text{PolyMult}}$.
The general case of the tree merge polynomial can be defined as

\vspace{-1em}
\begin{equation}\label{eq:Ftreemerge}
\mathcal{F}_{\text{TreeMerge}}(x,y;m,n)
:= \sum_{i=0}^{m-1}\Bigl(\prod_{j=0}^{n-1} (x_j\oplus y_j)^{E_{i,j}}\Bigr),
\end{equation}

\noindent where $E_{i,j}\in\mathbb{N}$ is the exponent of $(x_j\oplus y_j)$ in polynomial $i$, and $x$ and $y$ are the sender and receiver shares, respectively. In the baseline protocol of Figure~\ref{TEE_secure_multiplication}, the required correlated randomness (e.g., subset products of $r_j$) grows as

\vspace{-1em}
\begin{equation}\label{eq:Nnaive}
N_{\text{naive}}= \sum_{i=0}^{m-1}(2^{\sum_{j=0}^{n-1} E_{i,j}}-1).
\end{equation}

The first observation is that, viewed \emph{row-wise}, all shares in each $\mathcal{F}_{\text{PolyMult}}$ are Boolean. Hence, for bits $a,b\in\{0,1\}$ and any integer $n\ge 1$, $(a\oplus b)^n = a\oplus b$. Equivalently, for each term $(x_j \oplus y_j)^{E_{i,j}}$ the exponent does not change the value, i.e., $(x_j \oplus y_j)^{E_{i,j}} = x_j \oplus y_j$, and $N_{\text{naive}}$ simplifies to

\vspace{-1.0em}
\begin{equation}\label{eq:Nopt}
N_{\text{opt}}=\sum_{i=0}^{m-1}(2^{n-1}-1).
\end{equation}

The second observation is that, viewed \emph{column-wise}, multiple polynomials often contain the same factor \((x_j\oplus y_j)\), so the corresponding correlated randomness can be safely reused without weakening security. Let \(E\in\mathbb{N}^{m\times n}\) be the exponent matrix; a reuse opportunity exists whenever if column \(j\) has more than one positive entry \(E_{i,j}>0\). For each row \(i\), define the active index set $A_i \!:= \{j\in\{0,\ldots,n-1\}\!:\!E_{i,j}>0\}$. For any index subset \(T\subseteq\{0,\ldots,i-1\}\), write \(A_T := \bigcap_{t\in T} A_t\). The number of correlated randomness required is 

\begin{figure}[t]
  \centering
  \includegraphics[width=0.9\linewidth, trim={0cm 0cm 0cm 0cm}, clip]{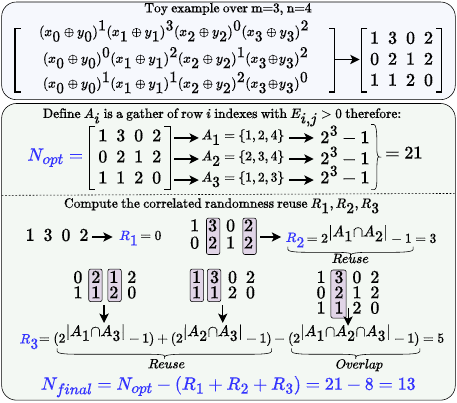}
  \vspace{0.03in}
  \caption{Toy example of correlated randomness reuse.}
  \label{polynomial_reuse}
\end{figure}

\vspace{-1em}
\begin{equation}\label{eq:Nfinal}
N_{final} = \sum_{\ell=0}^{i-1} (-1)^{\ell}
\sum_{\,T \in C^{\,i-1}_{\,\ell}}
\Bigl( 2^{\,\bigl|\,A_i \cap A_T \bigr|} - 1 \Bigr),
\end{equation}

\noindent where $C^{\,i-1}_{\,\ell}$ denotes all $\ell$-element subsets of $\{0,\ldots,i-1\}$. Intuitively, whenever two rows $i$ and $q$ share positions $j$ with $E_{i,j}>0$ and $E_{q,j}>0$, the associated randomness for $(x_j\oplus y_j)$ is reusable. 
A toy example is shown in Figure~\ref{polynomial_reuse}. The first row \(R_1\) has no reuse because it is the initial allocation. For the second row \(R_2\), reusable randomness is determined by the overlap of active sets \(A_1\) and \(A_2\), i.e., \(A_1 \cap A_2\). Similarly, for the third row \(R_3\), reuse is computed from the pairwise overlaps \(A_1 \cap A_3\) and \(A_2 \cap A_3\). Importantly, the triple overlap \(A_1 \cap A_2 \cap A_3\) must also be accounted for via inclusion--exclusion to avoid double subtraction when computing the total reusable randomness.

\section{TAMI-MPC Hardware Architecture}\label{sec:relu}

\begin{figure*}[t]
    \centering
    \includegraphics[width=\textwidth]{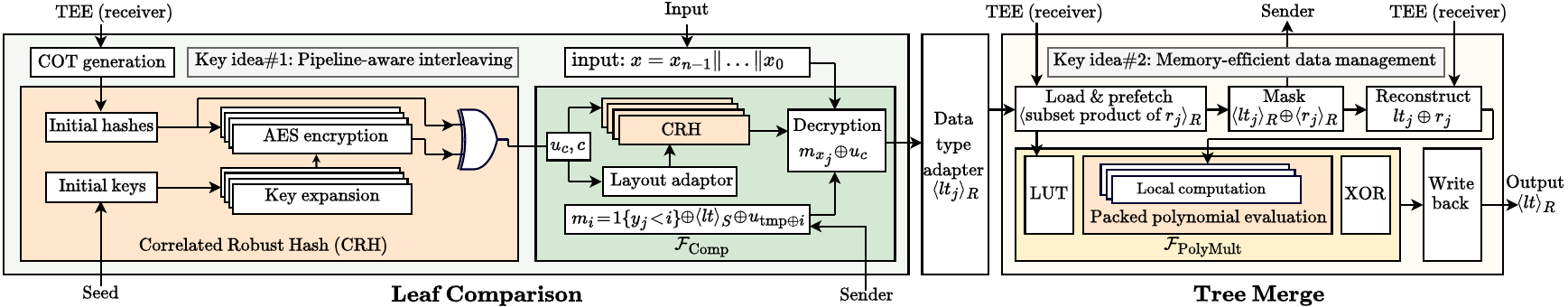}
    \caption{System-level accelerator architecture of TAMI-MPC.}
    \label{fig:overview}
\end{figure*}

In this section, we present an FPGA-based architecture design for TAMI-MPC's core primitives, TEE-assisted leaf comparison $\mathcal{F}_{\text{Comp}}$ and tree merge $\mathcal{F}_{\text{PolyMult}}$, as shown in Figure~\ref{fig:overview}.

\subsection{Overall Architecture} 

The leaf-comparison module mainly comprises a correlated OT (COT) generator~\cite{li2025silentflow} and a comparison unit ($\mathcal{F}_{\text{Comp}}$, Figure~\ref{TEE_secure_comparison}) whose compute core contains multiple subunits of correlation-robust hash (CRH)~\cite{Guo2020mitccrh}. The CRH derives the required ROT from the initial COT and prepares the mask used to decrypt the oblivious messages $m_i$ received from the sender, during online inference. The CRH operation and COT generation are the primary computational bottlenecks in leaf comparison; we mitigate it via pipeline-aware interleaving within each CRH core. The tree-merge module consists of two units, data exchange and polynomial evaluation ($\mathcal{F}_{\text{Polymult}}$, Figure~\ref{TEE_secure_multiplication}). The polynomial-evaluation stage incurs frequent data movement (e.g., reading pre-generated correlated randomness), which forms a major bottleneck; our memory-efficient data-management scheme alleviates this overhead.   

\subsection{Pipeline-Aware Interleaved Leaf Comparison $\mathcal{F}_{\text{Comp}}$}

\begin{figure}[t]
    \centering
    \includegraphics[width=0.95\linewidth, trim={0cm 0cm 0cm 0cm}, clip]{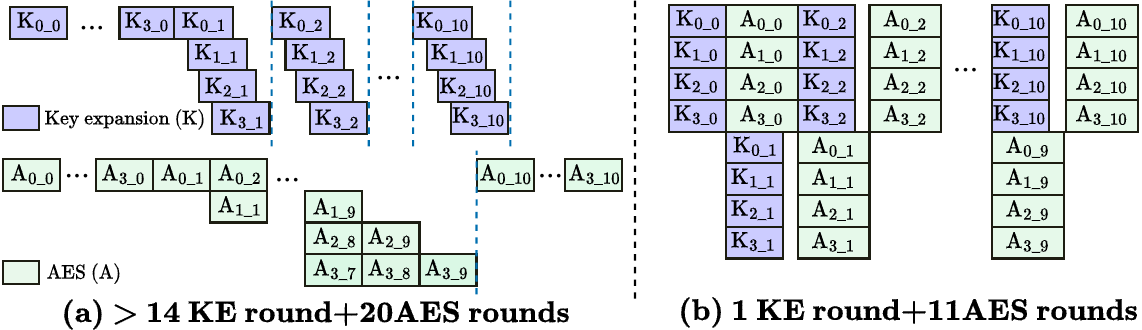}
    \caption{Pipeline demonstration of the hash function.}
    \label{fig:hash_pipe}
\end{figure}

CRH is used in OT to preserve the security of messages even when their inputs are correlated---for example, the message $m_i$ computed during leaf comparison. A typical AES-based CRH consists of two stages: seed-based key expansion and AES encryption. The key expansion stage derives a sequence of round keys from a seed, and the AES encryption applies the round function~\cite{dworkin2001advanced} to produce pseudorandom outputs that serve as the hash values. State-of-the-art works~\cite{rathee2020cryptflow2,huang2022cheetah} suffer from a loop-carried dependency induced by a read-modify-write pattern between these two stages, i.e., AES encryption must wait for key expansion to produce the corresponding round keys (Figure~\ref{fig:hash_pipe}(a)). Even worse, strictly sequential execution typically requires storing intermediate key-expansion results before they are consumed by the AES rounds, incurring additional memory traffic during the encryption phase.

To overcome these limitations, we propose an architecture that exploits \textit{data layout transformation} and \textit{spatial parallelism}. Instead of generating the key schedule for each AES block in sequence, our design interleaves the round keys of multiple independent blocks. This reorganization enables a deeply pipelined streaming architecture (Figure~\ref{fig:hash_pipe}(b)), allowing key expansion and encryption for multiple blocks to proceed concurrently. By eliminating the need to store intermediate key schedules, the proposed design achieves higher memory efficiency and substantially improves throughput. In our design, four parallel Key Expansion and AES units are instantiated to fully utilize the dataflow resources.

The theoretical advantages of this proposed architecture are summarized in Table~\ref{tab:prf_comparison}. The table provides an analytical comparison of computation cycles and memory transfers required to process $N$ data blocks, benchmarked against a conventional pipelined design. As shown, the proposed approach substantially reduces both computational complexity and memory traffic. The reduction in memory transfers, from $58N$ to only $6.25N$, is particularly significant, as it alleviates bandwidth bottlenecks by enabling local data reuse within the parallel processing units and fully utilizing the 512-bit AXI interface.

\begin{table}[h!]
    \centering
    \caption{Analytical comparison of conventional and proposed CRH evaluation architectures for processing $N$ blocks.}
    \label{tab:prf_comparison}
    \setlength{\tabcolsep}{2pt} 
    \footnotesize              
    \resizebox{0.95\columnwidth}{!}{%
    \begin{tabular}{@{}lll@{}}
        \toprule
        \textbf{Method} & \textbf{Computation Cycles} & \textbf{Memory Transfers} \\
        \midrule
        CPU      & $(11N+100)_{\text{KE}} + (11N+42)_{\text{AES}}$  & $22N_{\text{KE}} + 36N_{\text{AES}}$ \\
        Proposed & $max(13N/4_{\text{KE}}, 18N/4_{\text{AES}})$  & $12N/4_{\text{KE}} + 13N/4_{\text{AES}}$ \\
        \bottomrule
    \end{tabular}
    }
\end{table}


\subsection{Memory-Efficient Management for Tree Merge $\mathcal{F}_{\text{PolyMult}}$} 

The TEE-assisted tree merge typically comprises two stages: a one-round data exchange and local polynomial computation. During the exchange, the receiver send masked values \( \langle lt_j \rangle_R \oplus \langle r_j \rangle_R \) and locally reconstruct \(lt_j \oplus r_j \) via TEE (§\ref{sec:fusion}, Opt.\#1) for subsequent computation. These operations involve only lightweight XORs. However, the dominant bottleneck is data movement: buffering the exchanged values requires large on-chip storage, and fetching the correctly indexed correlated randomness triggers intensive memory reads. 

To address the memory bottleneck, we introduce a memory-efficient data-management scheme. Our key observation is that the access pattern for reading the corresponding correlated randomness is fixed across different comparisons and independent of the values \(lt_j\); only the randomness (e.g., the mask \( r_j \)) is updated, without branching on \(lt_j\), yielding a deterministic sequence of indices. This determinism enables a precomputed LUT, delivering a \(4.89\times\) speedup. Furthermore, by prefetching randomness and optimizing the dataflow in masked reconstruction, the masked data can be streamed directly to the final stage, reducing memory overhead and providing an additional \(5.56\times\) gain. 

We further reformulate the algorithm to exploit parallelism via a \textit{packed polynomial execution model}. The approach is driven by two observations: (1) the original data layout creates a stage imbalance---data exchange runs nearly four times faster than polynomial evaluation; and (2) the computations across leaf comparisons are independent and use only simple Boolean logic. We therefore encode multiple leaf comparisons into contiguous memory locations via the data type adaptor module (Figure~\ref{fig:overview}), enabling parallel polynomial evaluation followed by an XOR reduction.   
This fully packed layout exploits parallel evaluation of multiple comparisons per memory access, balances pipeline latency across stages, increases throughput, and improves memory efficiency. Consequently, up to $512/n$ comparisons ($n$ is the number of chunks during leaf comparison) and data transfers are completed per transmission, yielding an additional $154.84\times$ speedup.

\section{Experiments}\label{Sec:Simu}
\begin{table*}[htbp]
  \setlength{\tabcolsep}{2pt} 
  \centering
  \caption{Complexity comparison with state-of-the-art Millionaires' protocols. We use: \(n\) = number of chunks; \(k\) = input bitlength (bits); \(\lambda\) = security parameter (default \(\lambda=128\)). 
  }
  \resizebox{\textwidth}{!}{
    \setlength{\heavyrulewidth}{1.0pt}
    \begin{tabular}{m{15em}cccccccc} 
    \toprule
     
     \multirow{3}[3]{*}{\parbox{15em}{\centering \textbf{Schemes}}} & \multicolumn{4}{c}{\textbf{Leaf comparison ($\mathcal{F}_{\text{Comp}})$}} & \multicolumn{4}{c}{\textbf{Tree merge ($\mathcal{F}_{\text{PolyMult}})$}}\\ 
    
    \cmidrule{2-9}
    
    \multicolumn{1}{c}{} & \multicolumn{1}{c}{\centering Computation Operations} & \multicolumn{3}{c}{\centering Communication Cost (bits)} & \multicolumn{1}{c}{\centering Computation Operations} & \multicolumn{3}{c}{\centering Communication Cost (bits)} \\

    \cmidrule{2-9}
    
    \multicolumn{1}{c}{} & \multicolumn{1}{c}{\centering Offline} & \multicolumn{1}{c}{\centering Offline} & \multicolumn{1}{c}{\centering Online} & \multicolumn{1}{c}{\centering Round} & \multicolumn{1}{c}{\centering Offline} & \multicolumn{1}{c}{\centering Offline} & \multicolumn{1}{c}{\centering Online} & \multicolumn{1}{c}{\centering Round}\\
    
    \cmidrule{1-9} 

    \centering Cryptflow2~\cite{rathee2020cryptflow2}, BOLT~\cite{pang2024bolt} & $nk\times \binom{2}{1}\text{-ROT}$ (IKNP~\cite{kolesnikov2013improved})  & \( 2\lambda nk \) & \( n(k + 2^k) \) & 2 & $4(n-1)\times \binom{2}{1}\text{-ROT}$ & $8(n-1)(\lambda+1)$ & $8(n-1)$ & $\log_{2}n$\\ 
    \centering Cheetah~\cite{huang2022cheetah}, Bumblebee~\cite{lu2025bumblebee} &$nk\times \binom{2}{1}\text{-ROT}$ (Ferret~\cite{yang2020ferret}) & $\lambda^2(\log_{2}nk)/nk$ & \( n(k + 2^k) \) & 2 & $4(n-1)\times \binom{2}{1}\text{-ROT}$ & $4(n-1)$ & $8(n-1)$ & $\log_{2}n$\\ 
    \centering TAMI-MPC & $nk\times \binom{2}{1}\text{-ROT}$ (SilentFlow)~\cite{li2025silentflow}  & 0 & $nk$ & 1 & Eq.~\ref{eq:Nfinal} & 0 & $n-1$ & 1\\ 
    
    \bottomrule
    
    \end{tabular}%
  }
  \label{tab:complexity}%
\end{table*}%

\subsection{Experiment Setup}

We conduct our evaluation under three representative network settings: (1) a LAN scenario with 3\,Gbps bandwidth and 0.3\,ms latency, (2) a WAN-like configuration with 200\,Mbps bandwidth and 50\,ms latency, and (3) a mobile scenario with 100\,Mbps bandwidth and 80\,ms latency. For the secure computation primitives, we implement TAMI-MPC on a low-end Zynq-7000 SoC FPGA running at 170\,MHz, synthesized using Vitis High-Level Synthesis~\cite{vitis_hls_user_guide}. We define the client profile as a constrained configuration representative of typical IoT sensors (single-core 800MHz CPU, 512MB memory). In order to be able to run LLM (BERT-base) frameworks, we use a AMD 3995 64-core CPU as our edge server to reflect a realistic edge-AI resource profile.Intel SGX~\cite{costan2016intel} is used as the TEE testbed.

\subsection{Analysis of $\mathcal{F}_{\text{Comp}}$ and $\mathcal{F}_{\text{PolyMult}}$}\label{Sec:complexity_sec}

\begin{figure}[t]
    \centering
    \includegraphics[width=\linewidth, trim={0cm 0cm 0cm 0cm}, clip]{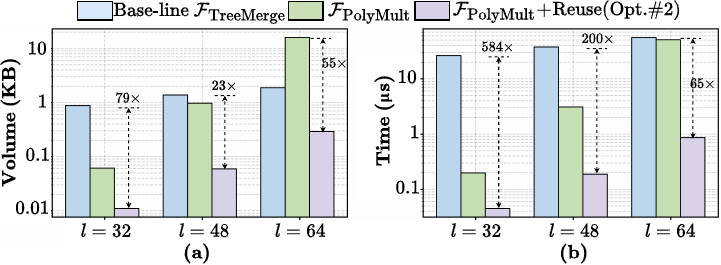}
    \caption{Correlated randomness generation comparison in tree merge (a) Volume (KB) (b) Time ($\mu$s). The y-axis is shown on a logarithmic scale.}
    \label{reuse}
\end{figure}

First, as shown in Table~\ref{tab:complexity}, the communication of TEE-assisted leaf-comparison drops from $n(k+2^k)$ to $nk$, and the number of interaction rounds reduces to one. Second, for tree merge, unlike prior approaches that generate ROT and then derive Beaver triples-based~\cite{lin2025ironman,feng2025panther} multiplication, our design produces the required shares directly inside the TEE, eliminating the ROT-to-triple operation and its well-known cache-bound overhead~\cite{li2025silentflow,lin2025ironman}. As a result, the offline phase is fully communication-free, and the online phase collapses from $\log_2 n$ to $1$ with one-eighth the communication volume under Opt.\#1 in §\ref{sec:fusion}. A potential concern with our one-round tree-merge protocol is that it appears to require exponential-level growth of correlated-randomness generation. However, with our reuse design in Opt.\#2 (§\ref{sec:fusion}), we eliminate the bulk of randomness-generation work, making its cost negligible. As shown in Figure~\ref{reuse}, evaluating secret input bitlengths from 32 to 64 bits yields \emph{two orders of magnitude} improvement—up to \textbf{$584\times$} speedup. 

\subsection{FPGA Acceleration}\label{sec:FPGA_result}

Table~\ref{tab:FPGA_MILL} summarizes the post-place-and-route hardware utilization and timing results running a data size of \(2\times 10^{5}\) for the CRH module, the leaf-comparison unit, the tree-merge stage, and the overall Millionaires' protocol during the online inference phase. We use Cheetah as our CPU baseline for comparison. Our pipeline-aware interleaving strategy achieves a $3.97\times$ speedup for the CRH module, which in turn enables a $2.72\times$ speedup in the leaf-comparison stage. Our memory-efficient data-management scheme reduces memory overhead and balances the performance--resource trade-off between stages, yielding a $80.15\times$ reduction in latency compared to Beaver's triple-based approach. Overall, the design uses about 30\% BRAM and 1\% DSP on a Z-7030, showing suitability for resource-limited embedded platforms.

\begin{table}[htbp]
\centering
\caption{Accelerator performance on the Zynq-7030 FPGA. Cheetah runs on an 800\,MHz CPU with 512\,MB of memory.}
\setlength{\tabcolsep}{2pt}      
\footnotesize                    
\resizebox{\columnwidth}{!}{
\begin{tabular}{lccccccc}
    \toprule
    \multirow{2}{*}{\textbf{Module}} &
    \multicolumn{4}{c}{\textbf{FPGA resources}} &
    \multicolumn{3}{c}{\textbf{Latency (ms)}} \\
    \cmidrule(lr){2-5} \cmidrule(lr){6-8}
    & BRAM & DSP & FF & LUT & Base & Ours & Spd. \\
    \midrule
    \multirow{2}{*}{CRH} 
    & 182 & 0 & 5082  & 13448 & 716.713 & -        & -          \\
    & 29  & 1 & 14352 & 12114 & -       & 180.451  & $3.97\times$ \\
    \midrule
    $\mathcal{F}_{\text{Comp}}$      & 15 & 2 & 28721 & 20679 & 486.496 & 179.205 & $2.72\times$ \\
    $\mathcal{F}_{\text{PolyMult}}$  & 21 & 0 & 10100 & 5515  & 117.017 & 1.46    & $80.15\times$ \\ 
    $\mathcal{F}_{\text{Mill}}$  & 38 & 2 & 41565 & 27758 & 608.271 & 179.909 & $3.38\times$ \\
    \bottomrule
\end{tabular}
}
\label{tab:FPGA_MILL}
\end{table}

\subsection{Nonlinear Evaluation}

\begin{figure}[t]
    \centering
    \includegraphics[width=\linewidth, trim={0cm 0cm 0cm 0cm}, clip]{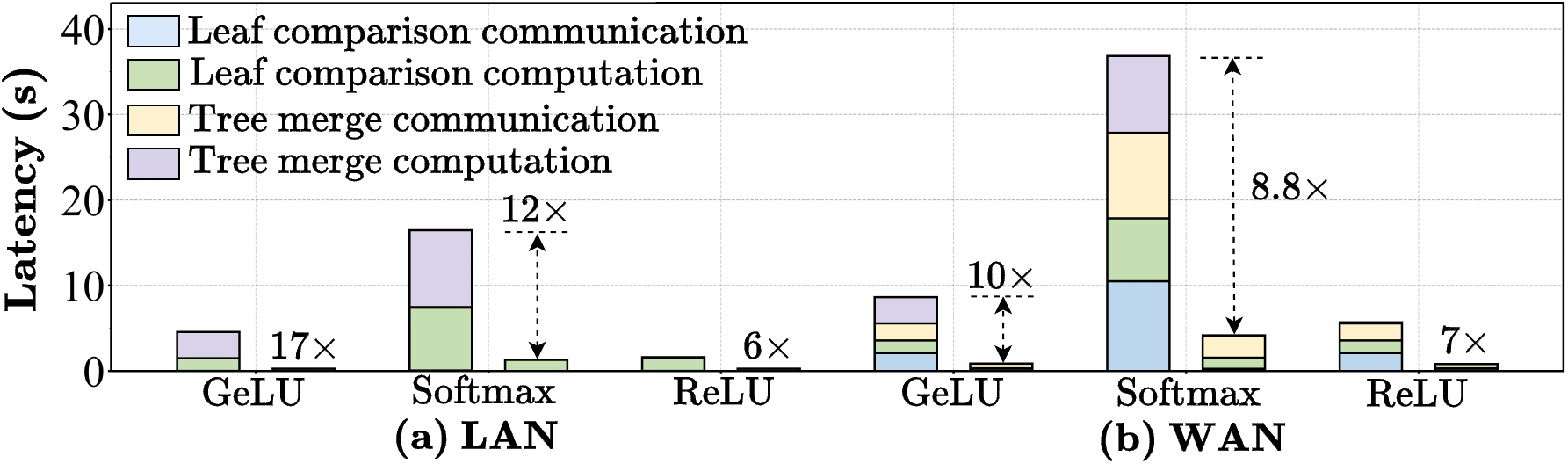}
    \caption{TAMI-MPC performance on different nonlinear activation layers. Input data size is \(2\times 10^{5}\).}
    \label{nonlinear}
\end{figure}

We evaluate TAMI-MPC on nonlinear activation microbenchmarks---ReLU, Softmax, and GeLU---summarized in Figure~\ref{nonlinear}. All three frameworks rely heavily on secure comparisons, which constitute the primary bottleneck. Beyond secure comparisons, the Softmax and GeLU layers are also approximated via polynomial evaluation (e.g., reciprocal, exponential)~\cite{lu2025bumblebee,pang2024bolt}. To compute these polynomials, we replace the original Beaver-triple–based polynomial multiplications with our $\mathcal{F}_{\text{Polymult}}$ to further optimize the computation. Our implementations follow Bumblebee’s Softmax and GeLU~\cite{lu2025bumblebee} and Cheetah’s ReLU~\cite{huang2022cheetah} protocol. With our design, the speedups are substantial: up to \(7\times\) over Cheetah’s ReLU in the WAN-network setting, and \(8.8\times\) and \(10\times\) over Bumblebee’s Softmax and GeLU, respectively. As network quality improves and communication overhead becomes negligible, the gains increase further, e.g., GeLU reaching up to \(17\times\) with FPGA acceleration.

\subsection{End-to-end Framework}

In Table~\ref{tab:framework_comparison}, we compare our framework against state-of-the-art PPMLaaS systems for CNNs and LLMs, including CryptFlow2, Cheetah, and Bumblebee. We evaluate three models: SqueezeNet (lightweight CNN), ResNet-50 (large-scale CNN), and BERT-base (lightweight LLM). Under a mobile network setting, our design completes inference in 108s for ResNet-50 and 24s for SqueezeNet, achieving \(4.86\times\)$\sim$\(4.95\times\) speedups over prior frameworks. For LLMs, we obtain a \(7.44\times\) speedup under the same setting. This improvement is even more pronounced than for CNNs, primarily because existing state-of-the-art frameworks rely heavily on approximating Softmax and GeLU via polynomial-based multiplication, which becomes a critical bottleneck; thus, nonlinear layer evaluation is far more dominant in LLM workloads. 
Overall, our design primarily targets nonlinear evaluation and demonstrates its effectiveness under resource-constrained devices and network conditions.

\begin{table}[htbp]
  \setlength{\tabcolsep}{2pt} 
  \centering
  \caption{Performance comparison with state-of-the-art CNN\&LLM PPMLaaS frameworks.}
  \footnotesize 
  \resizebox{\columnwidth}{!}{
    \setlength{\heavyrulewidth}{1.0pt}
    \begin{tabular}{m{5.5em}m{5em}ccccccccc} 
    
    \toprule
    
    \multirow{3}[-2]{*}{\parbox{5.5em}{\centering \textbf{Frame\\work}}} & \multirow{3}[-2]{*}{\parbox{5em}{\centering \textbf{Model}}} & \multicolumn{3}{c}{\textbf{LAN}} & \multicolumn{3}{c}{\textbf{WAN}}& \multicolumn{3}{c}{\textbf{Mobile}}\\
    
    \cmidrule{3-11}
    
    \multicolumn{1}{c}{} & \multicolumn{1}{c}{} &\multicolumn{1}{c}{\centering \shortstack{Base}} & \multicolumn{1}{c}{\centering \shortstack{Ours}} & \multicolumn{1}{c}{\centering \shortstack{Spd.}} & \multicolumn{1}{c}{\centering \shortstack{Base}} & \multicolumn{1}{c}{\centering \shortstack{Ours}} & \multicolumn{1}{c}{\centering \shortstack{Spd.}}& \multicolumn{1}{c}{\centering \shortstack{Base}} & \multicolumn{1}{c}{\centering \shortstack{Ours}} & \multicolumn{1}{c}{\centering \shortstack{Spd.}}\\
    
    \cmidrule{1-11} 
    
    \multirow{3}[-2]{*}{\parbox{5em}{\centering Cryptflow2~\cite{rathee2020cryptflow2}}} & SqueezeNet &335  & 178   &1.88$\times$& 550&182&3.02$\times$&821&206&3.98$\times$\\ 
    &ResNet-50&427  & 162   &2.63$\times$& 1034&292&3.54$\times$&1662&376&4.42$\times$\\ 
    
    \cmidrule{1-11}
    \multirow{3}[-2]{*}{\parbox{5em}{\centering Cheetah~\cite{huang2022cheetah}}} &SqueezeNet&101  & 23   &4.39$\times$&203&43&4.72$\times$&317&64&4.95$\times$\\ 
    &ResNet-50&168  &  88  &1.9$\times$& 381&97&3.92$\times$&525&108&4.86$\times$\\ 

    \cmidrule{1-11}
    
    \centering Bumblebee~\cite{lu2025bumblebee} &BERT-base& 701  &  272  &2.57$\times$& 1761&309&5.69$\times$&2828&380&7.44$\times$\\ 
    
    \bottomrule
    
    \end{tabular}%
  }
  \label{tab:framework_comparison}%
\end{table}%

\section{Conclusion}\label{Sec:Summary}

Secure MPC-based PPML inference is now within reach for resource-constrained platforms, as our design tackles the dominant performance bottlenecks in the underlying MPC primitives. TAMI-MPC employs a trusted-acceleration, minimally interactive MPC architecture that supports ResNet-50 inference in 108\,s over mobile networks on hardware comparable to IoT sensors or wearables, delivering a \(4.86\times\) speedup. For LLMs such as BERT-base, TAMI-MPC finishes inference under 380s, achieving a \(7.44\times\) speedup.

{\footnotesize 
\bibliographystyle{ACM-Reference-Format}
\bibliography{cite}

@String{Computing = "Computing" }

@String{Computer = "{IEEE} Computer" }

@String{Springer = "Springer-Verlag" }

@inproceedings{huang2022cheetah,
  title={Cheetah: Lean and fast secure Two Party deep neural network inference},
  author={Huang, Zhicong and Lu, Wen-jie and Hong, Cheng and Ding, Jiansheng},
  booktitle={31st USENIX Security Symposium (USENIX Security 22)},
  pages={809--826},
  year={2022}
}

@inproceedings{rathee2020cryptflow2,
  title={Cryptflow2: Practical 2-party secure inference},
  author={Rathee, Deevashwer and Rathee, Mayank and Kumar, Nishant and Chandran, Nishanth and Gupta, Divya and Rastogi, Aseem and Sharma, Rahul},
  booktitle={Proceedings of the 2020 ACM SIGSAC Conference on Computer and Communications Security (CCS)},
  pages={325--342},
  year={2020}
}

@inproceedings{pang2024bolt,
  title={Bolt: Privacy-preserving, accurate and efficient inference for transformers},
  author={Pang, Qi and Zhu, Jinhao and M{\"o}llering, Helen and Zheng, Wenting and Schneider, Thomas},
  booktitle={2024 IEEE Symposium on Security and Privacy (SP)},
  pages={4753--4771},
  year={2024},
  organization={IEEE}
}

@inproceedings{yang2020ferret,
  title={Ferret: Fast extension for correlated OT with small communication},
  author={Yang, Kang and Weng, Chenkai and Lan, Xiao and Zhang, Jiang and Wang, Xiao},
  booktitle={Proceedings of the 2020 ACM SIGSAC Conference on Computer and Communications Security (CCS)},
  pages={1607--1626},
  year={2020}
}

@inproceedings{zhou2022ppmlac,
  title={PPMLAC: high performance chipset architecture for secure multi-party computation},
  author={Zhou, Xing and Xu, Zhilei and Wang, Cong and Gao, Mingyu},
  booktitle={Proceedings of the 49th Annual International Symposium on Computer Architecture},
  pages={87--101},
  year={2022}
}

@article{huang2024efficient,
  title={Efficient Privacy-Preserving Machine Learning with Lightweight Trusted Hardware},
  author={Huang, Pengzhi and Hoang, Thang and Li, Yueying and Shi, Elaine and Suh, G Edward},
  journal={Proceedings on Privacy Enhancing Technologies},
  year={2024}
}

@inproceedings{kolesnikov2013improved,
  title={Improved OT extension for transferring short secrets},
  author={Kolesnikov, Vladimir and Kumaresan, Ranjit},
  booktitle={Advances in Cryptology--CRYPTO 2013: 33rd Annual Cryptology Conference, Santa Barbara, CA, USA, August 18-22, 2013. Proceedings, Part II},
  pages={54--70},
  year={2013},
  organization={Springer}
}

@article{lu2025bumblebee,
  title={Bumblebee: Secure two-party inference framework for large transformers},
  author={Lu, Wen-jie and Huang, Zhicong and Gu, Zhen and Li, Jingyu and Liu, Jian and Hong, Cheng and Ren, Kui and Wei, Tao and Chen, WenGuang},
  journal={Network and Distributed Systems Security (NDSS) Symposium},
  year={2025}
}

@article{zhang2021gala,
  title={GALA: Greedy computation for linear algebra in privacy-preserved neural networks},
  author={Zhang, Qiao and Xin, Chunsheng and Wu, Hongyi},
  journal={Network and Distributed Systems Security (NDSS) Symposium},
  year={2021}
}

@manual{vitis_hls_user_guide,
  title        = {Vitis High-Level Synthesis User Guide (UG1399)},
  author       = {{AMD}},
  year         = {2024},
  note         = {\url{https://docs.amd.com/r/en-US/ug1399-vitis-hls}},
}

@inproceedings{zhou2022efficient,
  title={Efficient privacy-preserving image classification for resource-constrained edge devices},
  author={Zhou, Yuxuan and Wang, Zihan and Wu, Lichao and others},
  booktitle={Proceedings of the 29th ACM Conference on Computer and Communications Security (CCS)},
  year={2022}
}

@inproceedings{juvekar2018gazelle,
  title={GAZELLE: A low latency framework for secure neural network inference},
  author={Juvekar, Chiraag and Vaikuntanathan, Vinod and Chandrakasan, Anantha},
  booktitle={27th USENIX security symposium (USENIX security 18)},
  pages={1651--1669},
  year={2018}
}

@inproceedings{rouhani2018deepsecure,
  title={Deepsecure: Scalable provably-secure deep learning},
  author={Rouhani, Bita Darvish and Riazi, M Sadegh and Koushanfar, Farinaz},
  booktitle={Proceedings of the 55th annual Design Automation Conference (DAC)},
  pages={1--6},
  year={2018}
}

@inproceedings{riazi2019xonn,
  title={XONN: oblivious deep neural network inference},
  author={Riazi, M Sadegh and Samragh, Mohammad and Chen, Hao and Laine, Kim and Lauter, Kristin and Koushanfar, Farinaz},
  booktitle={28th USENIX Security Symposium (USENIX Security 19)},
  pages={1501--1518},
  year={2019}
}

@inproceedings{zheng2019challenges,
    author = {Zheng, Mengyao and Xu, Dixing and Jiang, Linshan and Gu, Chaojie and Tan, Rui and Cheng, Peng},
    title = {Challenges of Privacy-Preserving Machine Learning in {IoT}},
    year = {2019},
    isbn = {9781450370134},
    publisher = {Association for Computing Machinery},
    address = {New York, NY, USA},
    doi = {10.1145/3363347.3363357},
    booktitle = {Proceedings of the First International Workshop on Challenges in Artificial Intelligence and Machine Learning for Internet of Things (AIChallengeIoT)},
    pages = {1–7},
    numpages = {7},
    location = {New York, NY, USA}
}

@article{aminifar2024privacy,
    title = {Privacy-preserving edge federated learning for intelligent mobile-health systems},
    journal = {Future Generation Computer Systems},
    volume = {161},
    pages = {625-637},
    year = {2024},
    issn = {0167-739X},
    doi = {https://doi.org/10.1016/j.future.2024.07.035},
    author = {Amin Aminifar and Matin Shokri and Amir Aminifar},
}

@inproceedings{mishra2020delphi,
  title={Delphi: A cryptographic inference system for neural networks},
  author={Mishra, Pratyush and Lehmkuhl, Ryan and Srinivasan, Akshayaram and Zheng, Wenting and Popa, Raluca Ada},
  booktitle={Proceedings of the 2020 Workshop on Privacy-Preserving Machine Learning in Practice},
  pages={27--30},
  year={2020}
}

@inproceedings{zhang2024individual,
  title={From Individual Computation to Allied Optimization: Remodeling Privacy-Preserving Neural Inference with Function Input Tuning},
  author={Zhang, Qiao and Xiang, Tao and Xin, Chunsheng and Wu, Hongyi},
  booktitle={2024 IEEE Symposium on Security and Privacy (SP)},
  pages={4810--4827},
  year={2024},
  organization={IEEE}
}

@article{costan2016intel,
  title={Intel SGX explained},
  author={Costan, Victor and Devadas, Srinivas},
  journal={Cryptology ePrint Archive},
  year={2016}
}

@article{feng2025panther,
  title={Panther: Practical secure 2-party neural network inference},
  author={Feng, Jun and Wu, Yefan and Sun, Hong and Zhang, Shunli and Liu, Debin},
  journal={IEEE Transactions on Information Forensics and Security},
  year={2025},
  publisher={IEEE}
}

@inproceedings{li2025silentflow,
  title={Silentflow: Leveraging Trusted Execution for Resource-Limited MPC via Hardware-Algorithm Co-design},
  author={Li, Zhuoran and Asl, Hanieh Totonchi and Nouri, Ebrahim and Cai, Yifei and Zhao, Danella},
  booktitle={31th Asia and South Pacific Design Automation Conference (ASP-DAC)},
  year={2026}
}

@article{lin2025ironman,
  title={Ironman: Accelerating Oblivious Transfer Extension for Privacy-Preserving AI with Near-Memory Processing},
  author={Lin, Chenqi and Yang, Kang and Xu, Tianshi and Liang, Ling and Wang, Yufei and Chen, Zhaohui and Wang, Runsheng and Gao, Mingyu and Li, Meng},
  journal={MICRO-58: 58th Annual IEEE/ACM International Symposium on Microarchitecture},
  year={2025}
}

@article{ryffel2022ariann,
  title={AriaNN: Low-Interaction Privacy-Preserving Deep Learning via Function Secret Sharing},
  author={Ryffel, Th{\'e}o and Tholoniat, Pierre and Pointcheval, David and Bach, Francis},
  journal={Proceedings on Privacy Enhancing Technologies},
  volume={1},
  pages={291--316},
  year={2022}
}

@article{maeng2024accelerating,
  title={Accelerating relu for mpc-based private inference with a communication-efficient sign estimation},
  author={Maeng, Kiwan and Suh, G Edward},
  journal={Proceedings of Machine Learning and Systems},
  volume={6},
  pages={128--147},
  year={2024}
}

@article{dworkin2001advanced,
  title={Advanced encryption standard (AES)},
  author={Dworkin, Morris J and Barker, Elaine and Nechvatal, James R and Foti, James and Bassham, Lawrence E and Roback, E and Dray Jr, James F and others},
  year={2001},
  publisher={National Institute of Standards and Technology (NIST), Morris J. Dworkin~…}
}

@inproceedings{Guo2020mitccrh,
    author = {Guo, Chun and Katz, Jonathan and Wang, Xiao and Weng, Chenkai and Yu, Yu},
    title = {Better Concrete Security for Half-Gates Garbling (in the Multi-instance Setting)},
    year = {2020},
    isbn = {978-3-030-56879-5},
    publisher = {Springer-Verlag},
    address = {Berlin, Heidelberg},
    url = {https://doi.org/10.1007/978-3-030-56880-1_28},
    doi = {10.1007/978-3-030-56880-1_28},
    booktitle = {Advances in Cryptology – CRYPTO 2020: 40th Annual International Cryptology Conference, CRYPTO 2020, Santa Barbara, CA, USA, August 17–21, 2020, Proceedings, Part II},
    pages = {793–822},
    numpages = {30},
    location = {Santa Barbara, CA, USA}
}
} 

\end{document}